\documentclass{aa}  

\usepackage{graphicx}
\usepackage{txfonts}
\usepackage{graphicx}	
\usepackage{amsmath}	
\usepackage{amsfonts}
\usepackage{multicol}   
\usepackage{bm}		    
\usepackage{pdflscape}
\usepackage{acro}
\usepackage[normalem]{ulem}
\usepackage{natbib}
\usepackage{hyperref}
\usepackage[capitalise]{cleveref}
\usepackage{xcolor}
\usepackage{placeins}
\usepackage{afterpage}
\usepackage{longtable}
\usepackage{multirow}
\usepackage{array}     
\usepackage{tabularx}   
\usepackage{makecell}

\usepackage{pgfplots}

\newcommand{\iap}{CNRS \& Sorbonne Universit\'{e}, Institut d’Astrophysique de Paris (IAP), UMR 7095, 98 bis bd Arago, F-75014 Paris, France}
\newcommand{\vienna}{Heuristic and Evolutionary Algorithms Laboratory, University of Applied Sciences Upper Austria, Hagenberg, Austria}
\newcommand{\oxford}{Astrophysics, University of Oxford, Denys Wilkinson Building, Keble Road, Oxford OX1 3RH, UK}
\newcommand{\icg}{Institute of Cosmology \& Gravitation, University of Portsmouth, Dennis Sciama Building, Portsmouth, PO1 3FX, UK}

\newcommand{\Mpch}{\ensuremath{h^{-1}\, \text{Mpc}}}
\newcommand{\hMpc}{\ensuremath{h\, \text{Mpc}^{-1}}}

\newcommand{\operon}{\textsc{operon}}

\newcommand{\bacco}{\textsc{bacco}}

\newcommand{\splitatcommas}[1]{%
  \begingroup
  \begingroup\lccode`~=`, \lowercase{\endgroup
    \edef~{\mathchar\the\mathcode`, \penalty0 \noexpand\hspace{0pt plus 1em}}%
  }\mathcode`,="8000 #1%
  \endgroup
}

\begin{document} 
   \title{syren-baryon: Analytic emulators for the impact of baryons on the matter power spectrum}
   \titlerunning{syren-baryon}

   \author{
    {Lukas Kammerer} \thanks{\href{mailto:Lukas.Kammerer@fh-hagenberg.at}{Lukas.Kammerer@fh-hagenberg.at}}\inst{1}
    \and
    Deaglan J. Bartlett \thanks{\href{mailto:deaglan.bartlett@physics.ox.ac.uk}{deaglan.bartlett@physics.ox.ac.uk}}\inst{2,3}
    \and
    {Gabriel Kronberger} \inst{1}
    \and
    {Harry Desmond} \inst{4}
    \and
    {Pedro G. Ferreira} \inst{3}
    }

   \institute{
        \vienna
        \and
        \iap
        \and
        \oxford
        \and
        \icg
}

   \date{Received XXX; accepted YYY}
 
  \abstract
   {Baryonic physics has a considerable impact on the distribution of matter in our Universe on scales probed by current and future cosmological surveys, acting as a key systematic in such analyses.}
   {We seek simple symbolic parametrisations for the impact of baryonic physics on the matter power spectrum for a range of physically motivated models, as a function of wavenumber, redshift, cosmology, and parameters controlling the baryonic feedback.}
   {We use symbolic regression to construct analytic approximations for the ratio of the matter power spectrum in the presence of baryons to that without such effects. We obtain separate functions of each of four distinct sub-grid prescriptions of baryonic physics from the CAMELS suite of hydrodynamical simulations (Astrid, IllustrisTNG, SIMBA and Swift-EAGLE) as well as for a baryonification algorithm. We also provide functions which describe the uncertainty on these predictions, due to both the stochastic nature of baryonic physics and the errors on our fits.}
   {The error on our approximations to the hydrodynamical simulations is comparable to the sample variance estimated through varying initial conditions, and our baryonification expression has a root mean squared error of better than one percent, although this increases on small scales.
   These errors are comparable to those of previous numerical emulators for these models.
   Our expressions are enforced to have the physically correct behaviour on large scales and at high redshift.
   Due to their analytic form, we are able to directly interpret the impact of varying cosmology and feedback parameters, and we can identify parameters which have little to no effect. 
   }
   {
   Each function is based on a different implementation of baryonic physics, and can therefore be used to discriminate between these models when applied to real data. 
   We provide publicly available code for all symbolic approximations found.
   }


   \maketitle
%

\section{Introduction}
\label{sec:introduction}

It is now clear that galaxy formation or, more generally, baryons, can play a significant role in the morphology and evolution of the large scale structure of the Universe \citep{Chisari_2018}. The intricate interplay between gravitational collapse and feedback processes due to supernova explosions and active galactic nuclei (AGN) can determine the amplitude of clustering on small to intermediate scales. This, in turn, means that, if one is to continue to use large scale structure as an accurate probe of fundamental physics, one must take into account the role of baryons and how their effects might contaminate the accuracy of cosmological constraints \citep{Amon_2022,Preston_2023}. Alternatively, one can view the statistics of the large scale structure of the Universe as new window onto the complex processes which are at play in galaxy formation.

While there have been significant steps in understanding galaxy formation and evolution, fuelled by observations from new observatories, it is still an open problem. From observations, there is not yet a consensus whether baryonic effects are large \citep{Hadzhiyska_2024,Bigwood_2024}, moderate \citep{Lu_2021,Chen_2023} or small \citep{Terasawa_2024,GarciaGarcia_2024}. From a theoretical point of view, the study of galaxy formation relies heavily on computer simulations which attempt to take into account a plethora of physical processes over a vast range of scales. Given the computational limitations, approximations have to be made on small scales, attempting to accurately model the effects of baryons on scales below the resolution of the simulation. The choices made for the ``sub-grid'' physics can play a crucial role in the outcome of cosmological simulations, with different choices (by different research groups) leading to significantly different outcomes \citep{Chisari_2018}. 

Our focus will be on the effect of baryons, and the large uncertainties in their modelling, on 2-point clustering statistics, specifically, the power spectrum of matter fluctuations. We define it as follows. Given a density of matter, $\rho_{\rm M}(t,{\bf x})$, at a time time, $t$ and spatial coordinates ${\bf x}$, we can define the density contrast, $\delta_{\rm M}$, through $\rho_{\rm M}={\bar \rho}_{\rm M}[1+\delta_{\rm M}(t,{\bf x})]$ where ${\bar \rho}_{\rm M}(t)$ is the mean matter density at time $t$. If, at a fixed time $t$, we take the spatial Fourier Transform of the density contrast, we obtain $\delta_{\rm M}({\bf k})$ (where, for ease of notation, the dependence on $t$ is implicit and ${\bf k}$ is the wavenumber). Assuming that the inhomogeneities in $\rho_{\rm M}$ are statistically homogeneous and isotropic, we can define the power spectrum, $P_{\rm M}(k)$, to be
\begin{eqnarray}
\langle \delta^*_{\rm M}({\bf k}') \delta_{\rm M}({\bf k})\rangle \equiv(2\pi)^3P(k)\delta^3({\bf k}-{\bf k}') ,\label{eq:pkdef}
\end{eqnarray}
where $\delta^3(\cdots)$ is the 3-D Dirac delta function. The shape and amplitude of the power spectrum, $P(k)$, characterise the large scale structure of the Universe and are sensitive to cosmological parameters such as the expansion rate, the fractional density of different components, the initial conditions, the mass of neutrinos and other such fundamental properties. It is its sensitivity to baryonic physics that we wish to characterise here.

There has been a considerable amount of work on the effect of baryons on clustering. Early on, baryon cooling \citep{White_2004} and the effects of hot inter cluster gas \citep{Zhan_2004} were shown to play an important role as was the impact of AGN feedback on $P(k)$ \citep{van_Daalen_2011}. The first attempts at quantifying the impact of baryons on the matter power spectrum from the OverWhelmingly Large Simulations (OWLS) suite \citep{Harnois_Deraps_2015} were then applied to data \citep{Kohlinger_2016,Kohlinger_2017,Hikage_2019}, showing that small scale information has a significant impact on both cosmological and baryonic constraints. In parallel, there has been an ongoing programme to incorporate the effect of baryons into models of clustering, either through the halo model \citep{Semboloni_2011,Fedeli_2014a,Fedeli_2014b,Mohammed_2014,Semboloni_2011,Semboloni_2013,Debackere_2020} and variants of it \citep{Mead_2015,Mead_2021,Lu_2021}, effective field theory methods \citep{Lewandowski_2015,Braganca_2021,Sullivan_2021}, through the process of "baryonification" \citep{Schneider_2015,Schneider_2019,Schneider_2020,Arico_2020} applied directly to $N$-body simulations and mock data, and through the use of machine learning  \citep{Troster_2019,Villaescusa-Navarro_2020}. More recently, more targetted simulations focussing on smaller scales have looked at the role black holes may play in baryonic processes 
\citep{Martin-Alvarez_2024}. While the inference of the impact of baryons on structure formation can be very sensitive to assumptions of a given hydrodynamical simulation \citep{CAMELS_presentation,Villanueva-Domingo_2022,Delgado_2023}, it has been shown that  reasonably robust observable quantities can be constructed \citep{Shao_2022,Wadekar_2023a,Villanueva-Domingo_2022_halo_mass,Villaescusa-Navarro_2021,Wadekar_2023b,Jo_2025}. 

A general trend has developed of constructing surrogates or {\it emulators} that can, efficiently and accurately, reproduce the impact of baryonic physics on large scale structure through the use of machine learning techniques. These involve the use of Principle Component Analysis \citep{Huang_2019,Eifler_2015,Foreman}, Gaussian Processes \citep{Sharma_2024,Giri_2021}, Random Forest methods \citep{Delgado_2023}, Enthalpy Gradient Descent schemes \citep{Dai_2018} and, most notably, Neural Networks \citep{Arico_2021_baryons}. In the latter case, it has been used to explore the impact of baryons on cosmological constraints from current data using small scales \citep{Arico:2023ocu,Garcia-Garcia:2024gzy}.

In this work, instead of leveraging these numerical machine learning techniques, we employ the supervised machine learning technique of symbolic regression \citep{Kronberger_2024}, which attempts to learn concise and accurate symbolic approximations directly from a dataset. This technique has been successfully applied to achieve percent-level accurate SYmbolic Regression ENhanced (\textsc{syren}) emulators of the matter power spectrum in the absence of baryons \citep{Bartlett_2024_linear,Bartlett_2024_syren,Sui_2024}, so this works extends these approximation to capture their effect; we name the set of models produced `\textsc{syren-baryon}.'

To characterise the effects of various models of baryonic feedback on the distribution of matter in the Universe, we wish to find symbolic approximations to the function
\begin{equation}
    \label{eq:S_definition}
    S (k, z, \bm{\theta}) \equiv \frac{P(k, z, \bm{\theta})}{P_{\rm nbody}(k, z, \bm{\theta})},
\end{equation}
where $k$ is the wavenumber, 
$z$ is the redshift, and $\bm{\theta}$ is a vector of cosmological and/or astrophysical parameters.
$P(k, z, \bm{\theta})$ is the matter power spectrum in the presence of baryons, and $P_{\rm nbody}(k, z, \bm{\theta})$ is the corresponding quantity for a ``$N$-body'' scenario, i.e. in the absence of all astrophysical processes where the only force is gravity.
\citet{Sharma_2024} showed that this is the only quantity required to model baryonic effects at the field level up to scales $k\sim 10 \,\hMpc$ due to the large cross-correlation between fields with and without astrophysical effects.

Through causality, baryons cannot have effects on the very largest scales, and thus we require that $S(k, z, \bm{\theta}) \to 1$ as $k \to 0$ for any $z$ and $\bm{\theta}$.
However, for other values of $k$, the shape of $S(k, z, \bm{\theta})$ is determined entirely by the model used to describe baryonic processes in the Universe, and can be very sensitive to this choice \citep{Chisari_2018}.
In this paper we consider broadly two classes of baryonic models: hydrodynamical simulations and baryonification algorithms, which are described in \cref{sec:hydro_sims,sec:baryonification}, respectively.
We consider four different hydrodynamical simulations and a single baryonification model, and thus derive five different functions for $S (k, z, \bm{\theta})$. 
We choose to emulate the baryonification model of \citet{Arico_2020}, although note that we could have equally chosen the \citet{Schneider_2015} model, which has a numerical emulator \citep{Giri_2021}.
By comparing our functions to observations, one could constrain the parameters of these models 
and, through Bayesian model comparison, determine which of these gives the most faithful representation of baryonic physics at the level of the power spectrum, given current data.

The paper is organised as follows. In \cref{sec:baryon_models} we describe the implementations of baryonic physics we consider and the different loss functions used for the hydrodynamical simulations versus the baryonification model. \cref{sec:symbolic_regression} describes the search for analytic approximations to $S(k, z, \bm{\theta})$ with symbolic regression and how we choose which model to select. These expressions are then presented in \cref{sec:results} and discussed in \cref{sec:discussion}. We conclude in \cref{sec:conclusion}. The main results of the paper are summarised in \cref{tab:fits,tab:sigma_fits}.
Throughout, we use ``$\log$'' to denote the natural logarithm, with base-10 logarithms written as ``$\log_{10}$.'' Whenever the wavenumber, $k$, appears in a mathematical expression, it is assumed to have units of $\hMpc$.

\section{Baryon models}
\label{sec:baryon_models}

\subsection{Hydrodynamical Simulations}
\label{sec:hydro_sims}

Our current understanding of galaxy formation and evolution has been improved greatly thanks to large-scale cosmological hydrodynamic simulations \citep{Dave_2019, Nelson_2019, Dubois_2014, Schaye_2015, Vogelsberger_2014, Lee_2021, Vogelsberger_2020, McCarthy_2017, Bassini_2020, Feng_2016, Dolag_2017}.
These simulations can reproduce galaxies that align well with observations in various aspects such as mass distributions, structures, and colours thanks to their detailed modelling of baryonic processes. 
However, due to the finite resolution of the simulations, many critical astrophysical mechanisms cannot be resolved, and thus one must rely on sub-grid models for processes such as star formation, the growth of supermassive black holes, stellar feedback, and AGN feedback.
These remain poorly understood, and therefore often require extensive tuning of parameters to match observational data \citep{Somerville_2015}, which poses limitations to their application in precision cosmological analyses.

Instead of having a single cosmological simulation with tuned hydrodynamical parameters, the Cosmology and Astrophysics with MachinE Learning Simulations (CAMELS) \citep{CAMELS_presentation,CAMELS_DR1,CAMELS_DR2} project provides a suite of hydrodynamical simulations with varying cosmological and astrophysical parameters.
This enables one to see the effect of such variation and is designed as a machine learning training set, so that one can either infer or marginalise over these parameters when applied to real data.

For this work, we focus on two sub-sets of the CAMELS suite, both of which are run from redshift $z=127$ to $z=0$, and are conducted in a periodic comoving volume of length $25 \, \Mpch$ with $256^3$ dark matter particles and, for those involving hydrodynamics, $256^3$ initial gas resolution elements.
Firstly, we consider the Latin Hypercube (LH) of simulations. These consist of 1000 simulations where the cosmological and astrophysical parameters are varied across a LH in the range given in \cref{tab:hydro_prior}. 
In particular, $\Omega_{\rm m}$ and $\sigma_8$ are allowed to vary across a large range, with all other cosmological parameters fixed to 
$\Omega_{\rm b} = 0.049$, $h = 0.6711$, $n_{\rm s} = 0.9624$, $M_{\rm \nu} = 0 {\rm \, eV}$, $w_0 = -1$, $w_a = 0$, $\Omega_{\rm K} = 0$.
The four astrophysical parameters which are allowed to vary ($\splitatcommas{A_{\rm SN1}, A_{\rm SN2}, A_{\rm AGN1}, A_{\rm AGN2}}$) are related to stellar and AGN
feedback, and are scaled such that a value of unity corresponds to the original implementation of the simulations considered. We use 80 \% of these simulations for training our models and the remaining 20 \% for testing.  

Given the uncertainty in how to model unresolved astrophysical processes, the CAMELS suite provides simulations with different sub-grid models. In this work, we consider the four models for which -- at the time of writing -- contain 1000 publicly available LH simulations and wish to find a different function for $S(k, z, \bm{\theta})$ for each prescription. These are:
\begin{enumerate}
    \item \textit{Astrid}: These simulations are run using the MP-Gadget code \citep{Springel_2021} with the Astrid sub-grid physics model \citep{Ni_2022,Bird_2012}.
    \item \textit{IllustrisTNG}: This suite utilises the Arepo code \citep{Springel_2010,Weinberger_2020} and employs the IllustrisTNG model of galaxy formation. This model is fully described in \citet{Weinberger_2017,Pillepich_2018}, where the data release for the fiducial simulation is provided in \citet{Nelson_2019}.
    \item \textit{SIMBA}: For these simulations, the SIMBA model \citep{Dave_2019} of galaxy formation is used, and these are run using the GIZMO code \citep{Hopkins_2015}.
    \item \textit{Swift-EAGLE}: This suite is an implementation of the EAGLE sub-grid model \citep{Schaye_2015,Crain_2015} within the Swift code \citep{Schaller_2024}.
\end{enumerate}
For a thorough description of these simulations, see the CAMELS papers \citep{CAMELS_presentation,CAMELS_DR1,CAMELS_DR2} and references therein.
It is important to note that the meaning of the baryonic physics parameters ($\splitatcommas{A_{\rm SN1}, A_{\rm SN2}, A_{\rm AGN1}, A_{\rm AGN2}}$) depends on the simulation considered, and therefore these cannot be directly compared across the simulations \citep[see Table 1 of][for a succinct summary of the meaning of these parameters in the different simulations]{Lovell_2024}.

\begin{figure}
    \centering
    \input{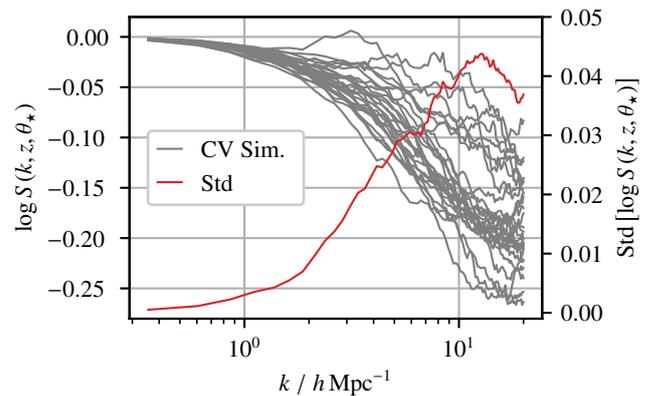}
    \caption{The baryonic effect, $\log S$ (\cref{eq:S_definition}), on the power spectrum for the 27 CV simulations and its standard deviation over $k$ for the Astrid sub-grid model at $z = 0$.
    }
    \label{fig:cv_set}
\end{figure}

As well as the LH set, for each sub-grid model and $N$-body code, there is a set of 27 simulations which take a fiducial set of cosmological and astrophysical parameters ($\Omega_{\rm m} = 0.3$, $\sigma_8 = 0.8$, $A_{\rm SN1} = A_{\rm SN2} = A_{\rm AGN1} = A_{\rm AGN2} = 1$), and sample the effect of cosmic variance (CV) using different initial random seeds. 
Although the cosmic variance for a pure $N$-body simulation would be small on these scales, different initial conditions can dramatically affect the astrophysical processes which occur, and thus a substantial variation in the power spectrum can be observed, particularly given the relatively small box sizes of the CAMELS simulations. This is depicted in \cref{fig:cv_set}, where we see that,
on large scales, all CV simulations predict the same power spectrum, but on small scales the effects of varying initial conditions and the stochastic nature of baryonic feedback can yield significantly different power spectra.

We use the CV simulations to estimate the sample variance of $S(k,z, \bm{\theta})$ by defining the variable 
\begin{equation}
    \sigma (k, z) = \max \left( 0.01, \underset{\rm CV}{\rm Std} \left[ \log S(k, z, \bm{\theta}_\star ) \right] \right),
\end{equation}
where ${\rm Std}$ denotes the standard deviation, which we compute over the CV simulations at fixed $k$ and $z$, and $\bm{\theta}_\star$ are the model parameters of the CV suite. We choose a lower bound of 0.01 since we are aiming for (at best) percent-level accuracy in our predictions, so are content with widening the error bars at values of $k$ for which the variation across the CV set is smaller than this.
$\sigma(k,z)$ has significant scale dependence, as demonstrated by the red line in \cref{fig:cv_set}. This is to be expected physically since, through causality, the very largest scales cannot be affected by feedback and the effects of cosmic variance on the dark-matter-only power spectrum have been divided out by computing $S$.
Similarly, $\sigma(k,z)$ is smaller at higher redshift since there as been less time for baryonic processes to impact the matter distribution, and thus all simulations have similar power spectra. 

One should expect that $\sigma$ depends on cosmological and astrophysical parameters, since, in the extreme case where there is no baryonic feedback, the simulations are identical to the $N$-body suite, and thus the value of $S$ would be unity independent of $k$, $z$ or the initial conditions of the simulation; the standard deviation over the CV set would then be zero.
However, since we only have a CV suite at the fiducial parameters, we are forced to assume that the errors are independent of these variables.
This is common practice in cosmological analyses, where covariance matrices are typically computed at a fiducial set of cosmological parameters~\citep{kodwani}.

Given the noisy estimate of $S(k,z,\bm{\theta})$ from a single simulation and its strong dependence on $k$, instead of using a mean squared error loss for our symbolic expressions, we minimise
\begin{equation}
    - \log \mathcal{L} [S_{\rm pred} ] = \sum_{i \in {\rm LH}} \sum_{k,z} \frac{\left( 
    \log S(k, z, \bm{\theta}_i) - \log S_{\rm pred}(k, z, \bm{\theta}_i)\right)^2}{2 \sigma^2 (k,z)},
\end{equation}
where $S(k, z, \bm{\theta}_i)$ is computed from simulation number $i$ on the LH, and $S_{\rm pred}(k, z, \bm{\theta}_i)$ is a candidate symbolic expression.
In fact, for numerical convenience, we do not directly optimise $\log \mathcal{L}$ but the normalised version
\begin{equation}
    \label{eq:NMSE_CAMELS}
    {\rm NMSE} \equiv \frac{\log \mathcal{L} [S_{\rm pred} ]}{\log \mathcal{L} [\bar{S}]},
\end{equation}
where we define $\overline{\log S}$ to be the mean of $ \log S$ over $k$, $z$ and $\bm{\theta}$.

We use 25 values of $k$, logarithmically spaced in the range $0.35-19.97 \, \hMpc$ and 12 redshift values $z \in [0.00, \allowbreak 0.05, \allowbreak0.10, 0.15, 0.21, 0.34, 0.54, 0.95, 1.48, 2.46, \allowbreak 4.01, \allowbreak 127.0]$.
To compute $S(k, z, \bm{\theta})$, we utilise the publicly available $P(k, z, \bm{\theta})$ and $P_{\rm nbody}(k, z, \bm{\theta})$, which were originally obtained using the \textsc{pylians} code \citep{Pylians}.
This loss function ensures that our expressions are better fitting at low $k$ and $z$ where the uncertainty from the simulation is lower.

\begin{table}
\centering
\caption{Range of the cosmological (upper) and hydrodynamical (lower) parameters used in the CAMELS simulations considered.}
\begin{tabular}{>{\centering\arraybackslash}m{2cm}|>{\centering\arraybackslash}m{1cm}>{\centering\arraybackslash}m{1cm}|>{\centering\arraybackslash}m{1cm}>{\centering\arraybackslash}m{1cm}}
  & \multicolumn{2}{c|}{Astrid } & \multicolumn{2}{c}{\thead{IllustrisTNG, SIMBA,\\Swift-EAGLE}} \\
  Parameter & Min. & Max. & Min. & Max.  \\
  \hline
  \hline
  $\Omega_{\rm m}$ & 0.1 & 0.5 & 0.1 & 0.5  \\
  $\sigma_8$       & 0.6 & 1.0 & 0.6 & 1.0  \\
  \hline
  $A_{\rm SN1}$    & 0.25 & 4 & 0.25 & 4  \\
  $A_{\rm SN2}$    & 0.5  & 2 & 0.5  & 2  \\
  $A_{\rm AGN1}$   & 0.25 & 4 & 0.25 & 4  \\
  $A_{\rm AGN2}$   & 0.25 & 4 & 0.5  & 2  \\
\end{tabular}
\label{tab:hydro_prior}
\end{table}

\subsection{Baryonification}
\label{sec:baryonification}

Instead of incorporating hydrodynamics into the simulation itself, an alternative approach is to first run a $N$-body, dark-matter-only simulation (which requires a much lower computational budget), and then augment the output to correct for this missing physics.
This approach -- ``baryonification'' \citep{Schneider_2015,Schneider_2019,Arico_2020} -- perturbs the final positions of the dark matter particles under a set physically motivated prescriptions for effects such as star formation, gas cooling and AGN feedback. 
As such, the power spectrum is modified due to the impact of baryons.

In particular, we wish to emulate this effect under the model of \citet{Arico_2020}.
In this case, the dark matter is assumed to quasi-adiabatically relax under the modification of the gravitational potential caused by the baryonic mass components.
Central galaxies are assumed to have mass profiles given by an exponentially-decaying power law which is fixed to $r^{-2}$.
Their masses are set by halo abundance matching.
These are surrounded by satellite galaxies of 20\% of the mass of the central, and are distributed according the dark matter profile. 
Gas is assumed to be bound in halos, with a distribution given by a double power law, where the transition and slopes are free parameters. 
Some of the gas is ejected, and is modelled as a constant density with an exponential cut-off with characteristic scale $\eta$.
The mass fraction of these two gas components is given by a parametric function of the host halo mass, and the total stellar and gas mass is designed to equal the cosmic density of baryons.

This model is described by 7 parameters, and we allow these to vary in the range given by \cref{tab:baryonification_prior}. The density profiles of hot gas in halos are given by $\theta_{\rm inn}$, $M_{\rm inn}$ and $\theta_{\rm out}$ with a mass fraction given by $M_{\rm c}$ and $\beta$. The extent of the ejected gas is given by $\eta$ and the characteristic halo mass scale for centrals is determined by $M_1$.
This prior range is chosen to match that of \citet{Arico_2021_baryons} and are similar to those of \citet{Arico_2021_simultaneous} since these are sufficiently wide to reproduce the clustering of several hydrodynamic simulations.
In \cref{tab:baryonification_prior} we also state the values chosen for these parameters in our fiducial case, which we will use to visualise the results of our symbolic fits.

\begin{table}
    \centering
    \caption{Ranges of the cosmological (upper) and baryonification (lower) parameters used and their fiducial values.}
    \begin{tabular}{c|c|c|c}
        Parameter & Minimum & Maximum & Fiducial \\
        \hline\hline
         $\sigma_8$ & 0.73 & 0.90 & 0.834 \\
         $\Omega_{\rm m}$ & 0.23 & 0.40 & 0.3175 \\
         $\Omega_{\rm b}$ & 0.04 & 0.06 & 0.049 \\
         $n_{\rm s}$ & 0.92 & 1.01 & 0.9624 \\
         $h$ & 0.60 & 0.80 & 0.6711 \\
         $M_{\rm \nu} / {\rm eV}$ & 0.0 & 0.4 & 0.0 \\
         $w_0$ & -1.15 & -0.85 & -1.0 \\
         $w_a$ & -0.3 & 0.3 & 0.0 \\
         \hline
         $\log_{10} M_{\rm c} / \left( h^{-1} \, {\rm M_\sun} \right)$ & 9.0 & 15.0 & 12.0 \\
         $\log_{10} \eta$ & -0.7 & 0.7 & -0.3 \\
         $\log_{10} \beta$ & -1.0 & 0.7 & -0.22 \\
         $\log_{10} M_1 / \left( h^{-1} \, {\rm M_\sun} \right)$ & 9 & 13 & 10.5 \\
         $\log_{10} M_{\rm inn} / \left( h^{-1} \, {\rm M_\sun} \right)$ & 9 & 13.5 & 13.4 \\
         $\log_{10} \theta_{\rm inn}$ & -2 & -0.5 & -0.86 \\
         $\log_{10} \theta_{\rm out}$ & -0.5 & 0 & -0.25 \\
    \end{tabular}
    \label{tab:baryonification_prior}
\end{table}

For computational convenience, instead of generating new suites of $N$-body simulations with the cosmological parameters drawn from the ranges given in \cref{tab:baryonification_prior} and then adjusting these using sampled values of the baryonification parameters, we utilise the \bacco{} emulator \citep{Arico_2021_baryons}. This emulator can reproduce $S(k, \bm{\theta})$ with percent-level precision across the desired parameter range and redshifts, and thus is sufficiently accurate for our purposes.

We generate two sets of 1000 simulations with 50 values of $k$ each for training and testing, and parameters drawn from a LH.
Values for $k$ are logarithmically spaced in the range $0.01-4.692 \, \hMpc$. Redshifts are uniformly sampled in the range of $0-1.5$ as the \bacco{} emulator does not support higher values.
Since we do not have access to a measure of uncertainty on these power spectra, we minimise the following normalised mean squared error
\begin{equation}
    {\rm NMSE} = \frac{\sum_{i \in {\rm LH}} \sum_k \left( \log S(k, \bm{\theta}_i) - \log S_{\rm pred}(k, \bm{\theta}_i)\right)^2}{\sum_{i \in {\rm LH}} \sum_k \left( \log S(k, \bm{\theta}_i) - \log \bar{S} \right)^2},
\end{equation}
which is identical to \cref{eq:NMSE_CAMELS} for constant $\sigma(k,z)$.

\section{Symbolic Regression}
\label{sec:symbolic_regression}

To find simple but accurate descriptions of baryonic effects on the power spectrum, we use the supervised machine learning technique of symbolic regression (SR) \citep{Kronberger_2024}.
We employ the genetic programming-based SR code \operon{} \citep{Burlacu_2020}, which utilises evolutionary processes inspired by natural selection, including crossover (breeding) and mutation, to develop an initial set of functions over successive generations. This method ensures that well-fitting and simple expressions are retained, while those that are poorly fitting or overly complex are eliminated. Consequently, after several iterations, a refined population of expressions that accurately approximate the dataset is produced. 
We selected \operon{} over other SR tools because of its superior performance in benchmark evaluations \citep{LaCava_2021,Burlacu_2023} and its proven effectiveness in cosmological and astrophysical research \citep{Bartlett_2024_linear,Bartlett_2024_syren,Sui_2024,AbdusSalam_2024}.

Free parameters in the expressions evaluated by \operon{} are optimised \citep{Kommenda_2020} using the Levenberg–Marquardt algorithm \citep{Levenberg_1944,Marquardt_1963}. This includes additional scaling parameters, which are present at each terminal node within the expression tree. The model length is defined as the number of nodes in an expression tree, excluding these scaling terms.
During the non-dominated sorting process, the objective values (the accuracy metric and model length relative to the maximum model length) of the equations are compared using $\epsilon$-dominance \citep{Laumanns_2002}. Here, a hyper-parameter $\epsilon$ is supplied to \operon, so that two objective values within an $\epsilon$ distance of each other are considered equivalent, which helps the algorithm converge. 
We choose $\epsilon = 10^{-7}$.
We only use arithmetic functions ($+, \times, \div, \mathrm{pow}$) and the chosen maximum model size (maximum model length of 60 and expression tree depth of 10) because we observed that non-linear functions in the function set, as well as allowing larger models, led to models with implausible structure or discontinuities within the given variable value ranges. We terminate the search after 1000 generations because \operon{} did not find any improvements after around 500 to 600 generations. We set the mutation rate (25\%), as well as the population size (1000) and selection procedure (tournament; group size=5) to their default values.

To choose a model, we consider the error on the test data set, the model complexity, and prior knowledge of the behaviour of our model at large scales and early times. 
In particular, we require two physically-motivated constraints:
\begin{enumerate}
    \item $S(k, z, \bm{\theta}) \to 1$ as $k \to 0$ for any $z$ and $\bm{\theta}$: On the largest scales, by causality arguments, baryons cannot have an effect, and thus the matter power spectrum must be equal to the dark-matter-only prediction.
    \item $S(k, z, \bm{\theta}) \to 1$ as $z \to \infty$ for any $k$ and $\bm{\theta}$: At high redshift, there has been insufficient time for baryonic processes to disrupt the matter distribution, and thus the matter power spectrum should be the same with or without baryons included.
\end{enumerate}

We calculate the Pareto front of the test error and the model length and select the model with the lowest error that analytically satisfies these constraints.
Since the structure of the chosen model occurs often multiple times in the final population with different parameters, we select the model structure with the parameters that minimize the description length \citep{Bartlett_2023} for hydrodynamical simulations.
For the baryonification model, we only use the NMSE as an error measure, as there is no uncertainty in the data. Given that $k$ starts at $0.35 \, \hMpc$, we add data for low $S(k=0.01\, \hMpc, z, \bm{\theta})$ for all $z$ and $\bm{\theta}$ in the training set of all the hydrodynamical simulations to guide the optimisation procedure toward solutions that satisfy the constraints in respect of $k$. For the hydrodynamical simulations, we correct any remaining bias on the training set over all values of $\bm{\theta}$ parameters with a separate function of $k$ and $z$, which is summed with our original model of $\log S(k, z, \bm{\theta})$. We approximate the bias again with \operon{} in the same way as the original function, but with a smaller tree size (maximum expression tree depth of 6 and maximum model length of 25) and a function set including exponentiation. 
We select the model with the lowest error that analytically satisfies the constraints from a Pareto front of model length and test error.

Not only do we wish to provide fits for $S(k, z, \bm{\theta})$, but also on the typical error on the prediction. This error arises due to an intrinsic scatter due to sample variance (\cref{fig:cv_set}), but also due to imperfect symbolic fits.
Including this theoretical covariance within cosmological analyses has been shown to reduce biases in cosmological parameters without requiring strict scale-cuts \citep{Maraio_2024}.
We model this error $\varepsilon$ (defined as half of the range of 68\% of the error at each $k$ and $z$) as a function of $k$ and $z$. 
Given the two physically motivated constraints for $S(k, z, \bm{\theta})$ are enforced to hold exactly, this yields the following corresponding constraints on the error
\begin{enumerate}
    \item $\varepsilon(k, z) \to 0$ as $k \to 0$.
    \item $\varepsilon(k, z) \to 0$ as $z \to \infty$.
\end{enumerate}
We use \operon{} to model $\varepsilon$ with the settings 
described above
but a maximum expression tree depth of 7, a maximum model length of 30 and a function set with exponentiation. Since the constraints are not enforced during \operon's search, we use them for model selection from \operon's Pareto front of model length and test error. We select the model with lowest error that satisfies these constraints. For all expressions, we repeat the search five times with different initial random seeds to ensure that it did not converge to local minima with insufficient accuracy, implausible structures, or constraint violations regarding $k$ and $z$.

\section{Results}
\label{sec:results}

After running \operon{}
for each of the baryonic physics models, we obtain the Pareto fronts plotted in \cref{fig:pareto}. For all of the models, we see that the training and validation losses are relatively similar, suggesting that our models have not over-fitted on the training set. For the hydrodynamical simulations, we find that the NMSE plateaus after a model length between 20 and 40 (depending on the subgrid model), whereas the loss for the baryonification scheme continues to decrease at higher lengths. We interpret this as due to the lack of stochasticity in the baryonification model, which allows us to obtain smaller NMSE than the noisier hydrodynamical simulation models. 

The vertical dashed lines in \cref{fig:pareto} indicate the chosen model lengths for each of the hydrodynamical models. These were chosen using the scheme described in \cref{sec:symbolic_regression}. The functional form of these equations are tabulated in \cref{tab:fits}, where we give an equation as a function of $k$ and $z$ in the first line of each row, then relate the free parameters of that equation to the cosmological and hydrodynamical parameters in the subsequent lines. In a cosmological analysis, one could either leave these parameters to be free and infer these, or, preferably, their dependence on the cosmological and hydrodynamical parameters could be used to constrain the strength of feedback for a given sub-grid model.

\begin{figure*}
\input{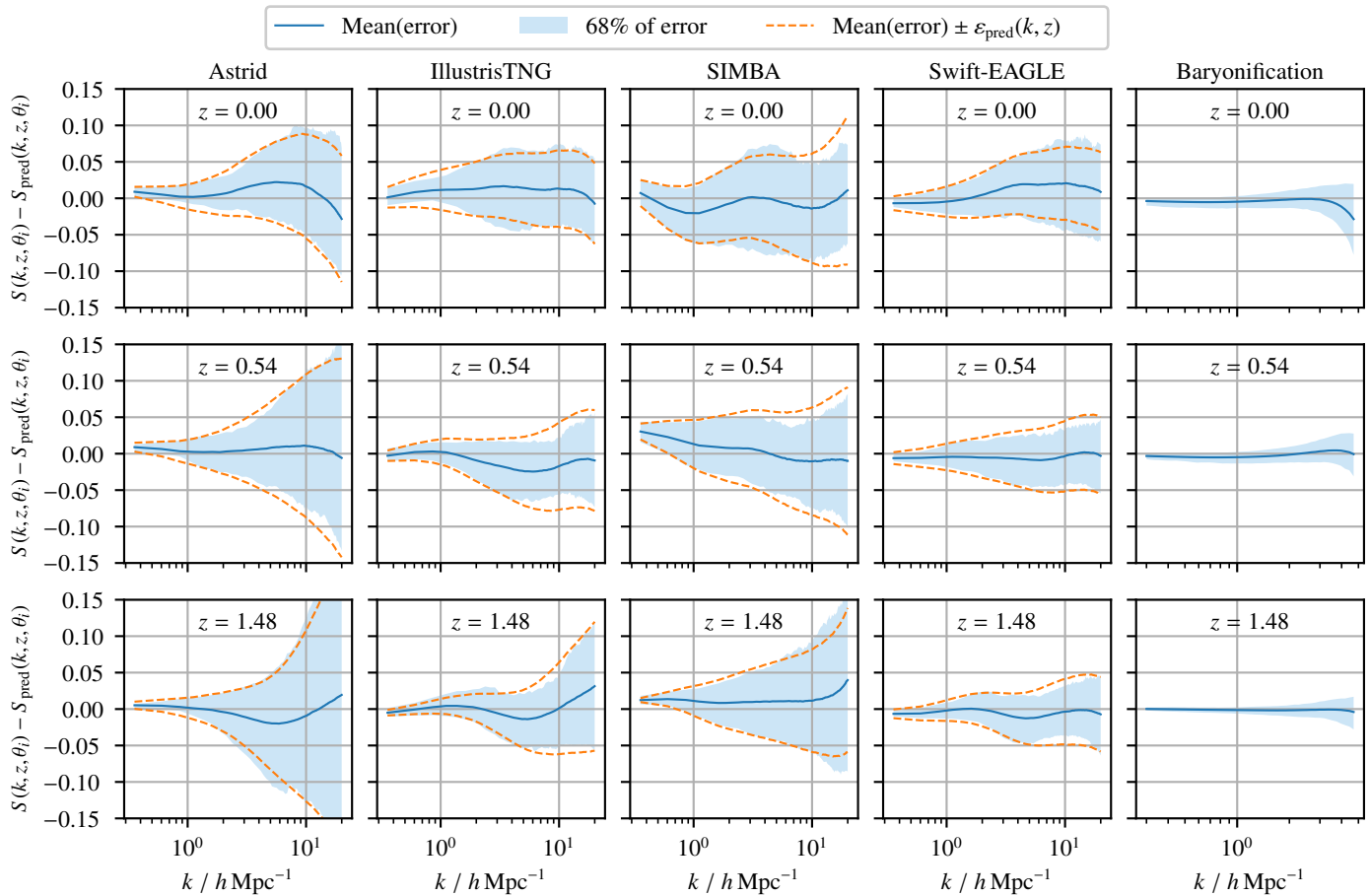}
\caption{
Difference between the symbolic fits for the baryonic effect on the power spectrum ($S_{\rm pred}$) to the true values ($S$) from the test set. 
Each column is for a difference baryonic model, and each row is for a different redshift, $z$. The solid lines indicate the mean 
difference and the shaded region contains 68\% of samples.
In the dashed orange lines we show the symbolic fits to the error on our models, as given in \cref{tab:sigma_fits}.
This error is due to a combination of stochasticity in the simulation and imperfect symbolic approximations.
}
\label{fig:error_on_S}
\end{figure*}

To determine the accuracy of these models, in \cref{fig:error_on_S} we plot the distribution of errors between our symbolic approximations to $S$ and the true values from the test set for a range of redshifts between $z=0$ and 1.5. We plot both the mean difference as a function of $k$ as well as the 68\% error band. Since we already include a term fit to remove the offset (although this is obtained using the training set), we see that the mean offset is approximately zero for all $k$, $z$ and model, and is consistent with zero within the 68\% error band.

For the baryonification model, the difference between the truth and prediction is strictly an error due to imperfect SR, and has a root mean squared value of 0.7\% for both the training and test sets, with increasing error as one increases $k$. 
This is the error on the fit to \bacco's emulation of the baryonification model, which matches the cosmological simulations to within 1-2\%. These errors should be combined to estimate the total error.

The hydrodynamical simulations, however, exhibit significant sample variance, and thus this error is a combination of the inherent stochasticity of the simulation and the error in our fit. To account for this, in \cref{fig:norm_error_on_S} we plot the normalised error on the test set, where we divide the difference between the truth and our prediction by the standard deviation over the CV set. If the CV set's standard deviation perfectly captured the stochasticity in the simulation and if our model was perfect up to this effect, the edges of the 68\% error bands in \cref{fig:norm_error_on_S} would be horizontal lines at $\pm 1$. This is not quite true, with the errors typically between 1.5 and 2 standard deviations, although for small $k$ we do achieve this ideal accuracy. Given the similar accuracy we obtain to \citet{Sharma_2024} (who use a flexible Gaussian Process to estimate $S$), we believe that this imperfection is dominated by the inaccuracy of the variance estimates from the CV set.

Given the significant stochasticity in the hydrodynamic models, for a cosmological analysis it is useful to have both a model for the mean behaviour but also this error, so that this theoretical uncertainty can be summed in quadrature with the covariance matrix of the likelihood. As such, we find symbolic approximations for the errors between the truth and predicted model, $\varepsilon$, as a function of $k$ and $z$. These are given in \cref{tab:sigma_fits}, and one can verify that these tend to zero for small $k$ and high $z$. These values are plotted in
\cref{fig:error_on_S},
where we see that these fits are able to smooth out the noisy estimates of this $\varepsilon$ due to the finite number of simulations used.

\begin{figure*}
\input{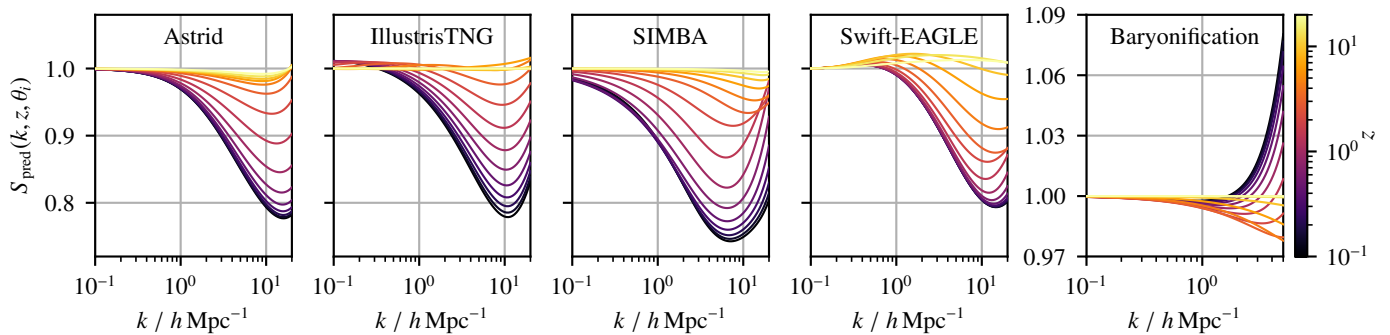}
\caption{
Extrapolation behaviour of our symbolic models for redshifts, $z$, and wavenumbers, $k$, outside of the range of the training data. The models are evaluated at the fiducial cosmological and astrophysical parameters (\cref{tab:baryonification_prior,tab:hydro_prior}). As required physically, the correction to the power spectrum, $S$, becomes unity at high redshift and on large scales (small $k$).
}
\label{fig:extrapolation}
\end{figure*}

One of the advantages of a symbolic emulator is the ability to enforce exact extrapolation behaviour in certain limits. As indicated in \cref{sec:symbolic_regression}, we wish to have models for which $S(k,z, \bm{\theta}) \to 1$ as $k \to 0$ or as $z \to \infty$. One can verify analytically that, for the prior range given in \cref{tab:hydro_prior,tab:baryonification_prior}, all of the models given in \cref{tab:fits} obey these requirements. To visualise these limits, in \cref{fig:extrapolation} we plot how our symbolic fits behave as a function of $k$ and $z$ as we take these parameters to very large or very small values, where we set the cosmological and hydrodynamical parameters to their fiducial values. One verifies that, indeed, $S$ becomes unity in the appropriate limits, and that this transition happens smoothly, with the magnitude of baryonic effects growing with time on small scales, whereas the largest scales remain relatively unaffected by their presence.

\begin{figure*}
\input{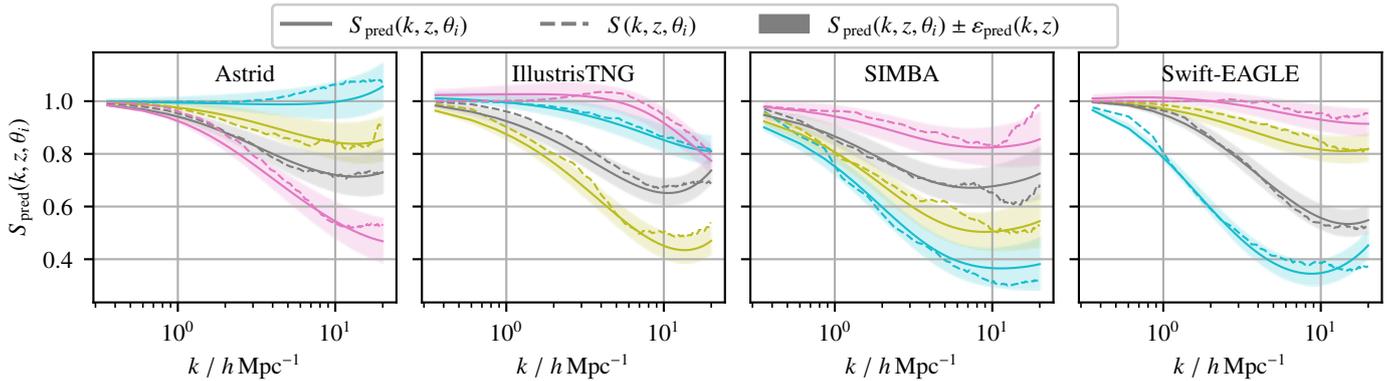}
\caption{
Predicted baryonic suppression of the matter power spectrum, $S$, as a function of wavenumber, $k$, at redshift zero for four randomly sampled hydrodynamical simulations at $z=0$. Our predictions from \cref{tab:fits} are shown as solid lines, and the shaded regions give the estimated error on these from \cref{tab:sigma_fits}. The dashed lines give the true values measured from the simulations, which are seen to be consistent with our predictions.
}
\label{fig:example_simulations}
\end{figure*}

Combining these results, in \cref{fig:example_simulations} we plot the true baryonic suppression for four randomly chosen simulations from each of the hydrodynamical simulations, alongside our predictions and error bands. One observes that our symbolic models are able to well-capture the broad features of this suppression, despite their different shapes and amplitudes as one varies the cosmological and hydrodynamical parameters. One sees the impact of sample variance, with noisy features in the ``true'' power spectra, which are smoothed into an estimate of the ensemble mean through our fits, yet whose variance is well-captured by $\varepsilon(k,z)$.

\begin{table*}
\centering
\caption{
    Chosen models for the baryonic correction to the power spectrum $\log S$ (\cref{eq:S_definition}) as a function of wavenumber $k$, redshift $z$ and the cosmological and hydrodynamical parameters outlined in \cref{tab:hydro_prior,tab:baryonification_prior}.
    }
\begin{tabular}{c|p{14cm}}
    Model & Fit for ${\log S(k, z, \bm{\theta})}$ \\
    \hline
    \hline \\
    Astrid & 
    {\vspace{-30pt}\begin{equation*}
    \frac{ \alpha_1 k \left(\alpha_{2} k\right)^{\alpha_{3}} \left(\alpha_{4} - z\right)}{\left(\alpha_{5} z - \alpha_{6}^{- 2 z}\right) \left(\alpha_{7} + \alpha_{8} k + k^{2}\right)} + \left(\alpha_{9} k - 1 + e^{- \alpha_{10} k}\right) e^{- \alpha_{11} z}
    \end{equation*}
    } \\
    & \vspace{-25pt} \begin{tabbing}
        \quad \= $\alpha_1 = 7.9 \frac{A_{\rm SN2}}{\Omega_{\rm m}}$, \quad \= $\alpha_{2} = 0.0014$, \quad \= $\alpha_{3} = 0.937 A_{SN2}$, \\ 
        \quad \= $\alpha_{4} = 1.622 A_{\rm AGN2} + 0.849 A_{\rm SN1} +  5.092 A_{\rm SN2}^2 \left( 0.23 A_{\rm AGN1} - A_{\rm AGN2} - 0.71 A_{\rm SN1} + 3.107 \sigma_{8} \right)$\\ 
        \quad \= $\alpha_{5} = 0.78$, \quad \= $\alpha_{6} = 0.677$, 
        \quad \= $\alpha_7 = A_{\rm AGN2} \left( 19.756 \Omega_{\rm m} + 2.8478 \right)$ \\
        \quad \= $\alpha_8 = 1.7347 A_{\rm AGN2} + 11.389 \Omega_{\rm m} + 1.642$ \quad \= $\alpha_{9} = 0.0224$, \quad \= $\alpha_{10} = 0.029$, \quad \= $\alpha_{11} = 0.063$
    \end{tabbing} \\
    \hline \\
    IllustrisTNG & 
    {\vspace{-30pt}\begin{equation*}
        - \alpha_{1} \alpha_{2}^{- \alpha_{3} k - \alpha_{4} z} k \left(1 + \alpha_{5} \left(- \alpha_{6} k + \alpha_{7}^{z}\right)\right) + \alpha_{8} e^{- \frac{\alpha_{9} e^{\alpha_{10} z}}{k}}
    \end{equation*}
    }
    \\
    & \vspace{-25pt} \begin{tabbing}
            \newline 
            \quad $\alpha_{1} = 0.0109 \frac{A_{\rm SN2}}{\Omega_{\rm m}}$, \quad \= $\alpha_{2} = 3592.322 \Omega_{\rm m}$, \quad \= $\alpha_{3} = 0.0087$, \quad \= $\alpha_{4} = 0.059$, \\
            \quad \= $\alpha_{5} = 0.9007 \frac{A_{\rm AGN2}}{A_{\rm SN2}} - 0.5901 \frac{A_{\rm SN1}}{A_{\rm SN2}} - 1.5576 + 2.6846 \frac{\sigma_8}{A_{\rm SN2}}$, \\ 
            \quad $\alpha_{6} = 0.048$, \quad \= $\alpha_{7} = \left(0.3 \Omega_{\rm m}\right)^{0.193}$, 
            \quad \= $\alpha_{8} = 0.022$, \quad \= $\alpha_{9} = 0.021$, \quad \= $\alpha_{10} = 0.797$
        \end{tabbing} \\
    \hline \\
    SIMBA & 
    {\vspace{-30pt}\begin{equation*}
        - \frac{ \alpha_{1} k \left(k + \alpha_{2}\right)}{\left( \left(\alpha_{3} k\right)^{\alpha_{4} z} + \alpha_{5} z\right) \left( k^{\alpha_{6}} + \alpha_{7} z + \alpha_{8}\right)} + \alpha_{9} e^{- \frac{\alpha_{10} e^{- \alpha_{11} z}}{k} - 2 z}
    \end{equation*}
    }
    \\
    & \vspace{-25pt} \begin{tabbing}
            \quad $\alpha_{1} = \frac{0.00133}{\Omega_{\rm m}^{2}}$, \quad \= $\alpha_{2} = \Omega_{\rm m} \left(25.727 A_{\rm AGN1} + 153.382 A_{\rm AGN2} - 70.6363 A_{\rm SN1} + 260.812 \sigma_{8}\right)$, \\
            \quad \= $\alpha_{3} = 0.0553$, \quad $\alpha_{4} = 0.79055$, \quad \= $\alpha_{5} = 0.6024$, \quad \= $\alpha_{6} = 1.594$, \quad \= $\alpha_{7} = 16.01$, \\
            \quad $\alpha_{8} = 12.0685 \left(1.0047 A_{\rm SN2}\right)^{0.6788 A_{\rm SN1}}$, \quad \= $\alpha_{9} = 4.4661$, \quad \= $\alpha_{10} = 114.529$, \quad \= $\alpha_{11} = 1.21$
        \end{tabbing} \\
    \hline \\
    Swift-EAGLE & 
    {\vspace{-30pt}\begin{equation*}
        \alpha_{1}^{\frac{\alpha_{2}}{k} + \alpha_{3} k + \alpha_{4} z} \left(- \alpha_{5} k + \alpha_{6} z + \frac{\alpha_{7} k z + k^{2}}{\alpha_{8} k + \alpha_{9}} + \frac{\alpha_{10} k + \alpha_{11}}{\alpha_{12} z + \alpha_{13} + k} + \alpha_{14}\right) + \alpha_{15} e^{- \frac{\alpha_{16} e^{- z}}{k} - e^{\alpha_{17} z}}
    \end{equation*}
    }
    \\
    & \vspace{-25pt} \begin{tabbing}
            \quad $\alpha_{1} = 0.272 \Omega_{\rm m}$, \quad \= $\alpha_{2} = 0.22$, \quad \= $\alpha_{3} = 0.0168$, \quad \= $\alpha_{4} = 0.074$, \quad \= $\alpha_{5} = \frac{0.0204}{\Omega_{\rm m}^{1.625}}$, \quad \= $\alpha_{6} = \frac{0.004}{\Omega_{\rm m}^{1.625}}$, \\
            \quad \= $\alpha_{7} = 1.1166$, \quad \= $\alpha_{8} = 54.3014 \Omega_{\rm m}^{1.625}$, \quad \= $\alpha_{9} = 0.12055 \Omega_{\rm m}^{1.625} \left(4822.2 A_{\rm SN1} + 8228.9 \Omega_{\rm m}\right)$, \\
            \quad \= $\alpha_{10} = \frac{- 0.2588 A_{\rm AGN1} + 6.13845 A_{\rm SN1}}{\Omega_{m}^{1.625} \left(6.528 A_{\rm AGN2} + 12.855 A_{\rm SN1}\right)}$, \quad \= $\alpha_{11} = \frac{\left(0.9788 - 0.1659 A_{\rm SN1}\right) \left(- 0.3745 A_{\rm AGN1} + 8.877 A_{\rm SN1}\right)}{\Omega_{\rm m}^{1.625} \left(6.528 A_{\rm AGN2} + 12.855 A_{\rm SN1}\right)}$,
            \quad \= $\alpha_{12} = 9.2437$, \\ \quad \= $\alpha_{13} = 9.2965 A_{\rm SN2} + 18.4$, \quad \= $\alpha_{14} = \frac{0.034 \Omega_{\rm m} - 0.02 \sigma_{8}}{\Omega_{\rm m}^{1.625}}$, \quad \= $\alpha_{15} = 2.287$, 
            \quad \= $\alpha_{16} = 130.0$, \quad \= $\alpha_{17} = 0.565$,
       \end{tabbing} \\
    \hline \\
    Baryonification  & 
    {\vspace{-30pt}\begin{equation*}
        \alpha_{1} \alpha_{2}^{\alpha_{3} z} k \left(\alpha_{4} - \left(\alpha_{5} k\right)^{\alpha_{6}}\right) \left(\alpha_{7} k + \alpha_{8}^{\left(\alpha_{9} + \alpha_{10} z\right)^{\alpha_{11}}}\right)
    \end{equation*}
    }
    \\
    & \vspace{-25pt} \begin{tabbing}
            \quad $\alpha_{1} = 2.4663 \left(5.823 \Omega_{\rm b}\right)^{3.23 \Omega_{\rm m}}$, \quad \= $\alpha_{2} = 5.823 \Omega_{\rm b}$, \quad \= $\alpha_{3} = 0.4355$, \\
            \quad \= $\alpha_{4} = 0.0495 \log_{10}(M_1)$, \quad \= $\alpha_{5} = 0.2479$, \quad \= $\alpha_{6} = - 0.27 \log_{10}(\eta)$, \\
            \quad \= $\alpha_{7} = 0.0015 \log_{10}(M_{\rm inn}) + 0.0176 \log_{10}(\theta_{\rm inn}) - 0.0015 \left(3.9144 \sigma_{8}\right)^{0.2058 \log_{10}(M_1)}$, \\
            \quad \= $\alpha_{8} = 0.016 \log_{10}(M_c)$, \quad \= $\alpha_{9} = 0.0706 \log_{10}(M_c)$, \quad \= $\alpha_{10} = 0.0383$, \quad \= $\alpha_{11} = - 6.5 \log_{10}(\beta) - 6.5$
    \end{tabbing}
\end{tabular}
    \label{tab:fits}
\end{table*}

\begin{table}
\centering
\caption{Chosen model for the error on baryonic correction to the power spectrum as a function of wavenumber $k$ and redshift $z$.}
\begin{tabular}{c|p{5cm}}
    Model & Fit for $\varepsilon(k, z)$ \\
    \hline
    \hline \\
    Astrid & 
    {\vspace{-30pt}\begin{equation*}
        \frac{\alpha_{1} k}{\alpha_{2} k e^{- \alpha_{3} z} + e^{\alpha_{4} z}}
    \end{equation*}
    } 
    \vspace{-10pt}
    \\
    & $\alpha = \{ \splitatcommas{0.0202, 0.1833, 1.3, 0.2529} \}$\\
    \hline \\
    IllustrisTNG & 
    {\vspace{-30pt}\begin{equation*}
        \frac{k}{\alpha_{1} k e^{- \alpha_{2} z} + \alpha_{3} z + \alpha_{4}}
    \end{equation*}
    } 
    \vspace{-10pt}
    \\
    & $\alpha = \{ \splitatcommas{17.119, 0.63, 48.797, 19.238} \}$\\
    \hline \\
    SIMBA & 
    {\vspace{-30pt}\begin{equation*}
        \alpha_{1} e^{- \frac{\alpha_{2} e^{\alpha_{3} z}}{k} + \alpha_{4} k}
    \end{equation*}
    } 
    \vspace{-10pt}
    \\
    & $\alpha = \{ \splitatcommas{0.059, 0.43, 0.6324, 0.0285} \}$\\
    \hline \\
    Swift-EAGLE & 
    {\vspace{-30pt}\begin{equation*}
        \frac{\alpha_{1} k}{\alpha_{2} k + e^{\alpha_{3} z}}
    \end{equation*}
    } 
    \vspace{-10pt}
    \\
    & $\alpha = \{ \splitatcommas{0.032, 0.54, 0.363} \}$\\
\end{tabular}
    \label{tab:sigma_fits}
\end{table}

\section{Discussion}
\label{sec:discussion}

\subsection{Dependence on cosmological and hydrodynamic parameters}
\label{sec:parameter_dependence}

One advantage of obtaining analytic expressions for $S$ is that we can now easily interpret the impact of varying individual cosmological and hydrodynamical parameters. In this subsection, we investigate this behaviour, beginning with the impact of cosmology in \cref{sec:cosmo_dependence}, before turning to hydrodynamics in \cref{sec:hydro_dependence}.
For the feedback parameters, we find that changing these values can give rise to opposite effects when comparing two hydrodynamical simulations, justifying our approach of conditioning on the model rather than trying to find a single symbolic approximation to fit all these simulations.

With such a comparison, there is a risk that the effects described below are peculiarities of the particular chosen model, rather than an intrinsic property of the hydrodynamic model. To verify this, we performed an additional five \operon{} runs per model and picked one Pareto-optimal expression from each run with similar or higher complexity and accuracy as the one given in \cref{tab:fits}. 
For each of these functions, we verified that the variation with cosmological and hydrodynamical parameters is similar to ensure the robustness of our qualitative conclusions.

\subsubsection{Cosmological dependence}
\label{sec:cosmo_dependence}

The impact of cosmology for the SIMBA and Astrid models is particularly straightforward. For SIMBA, one sees that the only parameters affected by cosmology are $\alpha_1 \propto \Omega_{\rm m}^{-2}$ and $\alpha_2 \propto \Omega_{\rm m} (f(A) + \sigma_8)$, where $f(A)$ is a function of the hydrodynamic parameters. Hence, one finds that, at large $k$, $\log S \propto \Omega_{\rm m}^{-2}$, and at small $k$ we have $\log S \propto \Omega_{\rm m}^{-1} (f(A) + \sigma_8)$. Similarly, for the Astrid model, $\alpha_1 \propto \Omega_{\rm m}^{-1}$, whereas $\alpha_7$ and $\alpha_8$ are linear in $\Omega_{\rm m}$, and $\alpha_4$ is linear in $\sigma_8$. These combinations mean that the power suppression decreases as $\Omega_{\rm m}$ increases for both of these models, which is consistent with \citet{Delgado_2023}'s analysis of the SIMBA model. One difference we find with \citet{Delgado_2023} is the dependence on $\sigma_8$; \citet{Delgado_2023} suggests that varying $\sigma_8$ with all other parameters fixed does not 
drive systematic variations in $S$, whereas we find that, for both models, increasing $\sigma_8$ increases the effect of baryons.

The IllustrisTNG model has slightly more complex cosmological behaviour. The $\sigma_8$ dependence is simple to read off, since $\alpha_5$ is linear in $\sigma_8$ and thus increasing $\sigma_8$ increases the effect of baryons. The impact is weaker than varying $\Omega_{\rm m}$, as noted by \citet{Delgado_2023}, who also studied this model. We see that $\alpha_1 \propto \Omega_{\rm m}^{-1}$ and that $\alpha_2 \propto \Omega_{\rm m}$ (but $\alpha_2$ is raised to a negative power), hence the magnitude of the prefactor of the first term in our fit is a decreasing function of $\Omega_{\rm m}$. However, we also have the term $\alpha_7^z \propto \Omega_{\rm m}^{0.193 z}$ to consider. For $z=0$ this has no effect, and thus we arrive at the same conclusion as \citet{Delgado_2023} that increasing $\Omega_{\rm m}$ decreases the effect of baryons on the matter power spectrum. Indeed, provided $z < 1 / (1 - \alpha_4) \approx 1.06$, this term cannot result in the opposite trend. For small $k$, one can ignore the $k$ dependence inside the bracket of the first term of our expression, and thus consider the function $1 + \alpha_5 \alpha_7^z$. For the fiducial feedback parameters at $z \sim 1$, $\alpha_5 \alpha_7^z \lesssim 0.62$, and thus this term does not dominate, resulting in the same $\Omega_{\rm m}$ behaviour as before. However, for large values of $\alpha_5$ (corresponding to large values of $A_{\rm AGN2} / A_{\rm SN2}$), one finds that this is no longer true and thus the effect of baryons on the power spectrum can be positively correlated with $\Omega_{\rm m}$ provided AGN feedback is much stronger than supernova feedback and at high redshift.

Swift-EAGLE also exhibits a slightly complex dependence on $\Omega_{\rm m}$. 
Most terms inside the bracket of the first term are proportional to $\Omega_{\rm m}^{-1.625}$, however $\alpha_{14}$ also contains a term proportional to $\Omega_{\rm m}^{-0.625}$. The complication arises due to the exponent $\alpha_1$ is raised to, since $\alpha_1 \propto \Omega_{\rm m}$. For very large and small $k$, this exponent will become larger than $0.625$, and is thus guaranteed to lead to a positive correlation between the magnitude of $\log S$ and $\Omega_{\rm m}$. However, at $z=0$ this only occurs for $k < 0.36 \, \hMpc$ and $k > 37 \, \hMpc$, which is outside the range of our training set and thus is an extrapolation effect. Hence, within the range of the training data at $z=0$, increasing $\Omega_{\rm m}$ decreases the effect of baryons on the power spectrum. At $z=1$, some of the small $k$ modes do have the opposite trend, but this effect is negligible given that baryonic suppression is small on these scales.

For all the cosmological simulations, only $\Omega_{\rm m}$ and $\sigma_8$ were varied, so we could not determine the effect of other parameters. However, for the baryonification model we varied many more parameters so can study their impact. Despite this, we find that many cosmological parameters do not appear in our expressions ($h$, $n_{\rm s}$, $M_{\rm \nu}$, $w_0$ and $w_a$), and thus their effect on the baryonic suppression of the power spectrum is minimal (this is also noted by \citet{Arico_2021_baryons}). Interestingly, $\Omega_{\rm b}$ does appear, since having fixed $\Omega_{\rm b}$ is the reason given in \citet{Delgado_2023} for the impact of $\Omega_{\rm m}$; as $\Omega_{\rm m}$ decreases the baryon fraction of the total matter content increases, making the process of expelling gas out of halos more efficient. 
\citet{Arico_2021_baryons} also found the dependence on $\Omega_{\rm b}$ to be important, in particular in the combination $\Omega_{\rm b}  / \Omega_{\rm m}$.
We see that the effect of baryons within the baryonification model is positively correlated with $\Omega_{\rm b}$, and find the same $\Omega_{\rm m}$ dependence as before; for fixed $\Omega_{\rm b}$ (and given that $5.823\Omega_{\rm b} < 1$ for our prior), increasing $\Omega_{\rm m}$ decreases the magnitude of $\log S$. $\sigma_8$ also appears in these expressions with the same trend as before for our prior range.

\subsubsection{Dependence on feedback parameters}
\label{sec:hydro_dependence}

We now consider the effect of the parameters controlling baryonic physics on the matter power spectrum. 
Starting with the SIMBA model, we see that both AGN parameters appear linearly in the $\alpha_2$ term of $\log S$, suggesting that increasing either parameter will increase the suppression of power. Moreover, the coefficient of $A_{\rm AGN2}$ is much larger than for $A_{\rm AGN1}$, hence $\log S$ is far more sensitive to $A_{\rm AGN2}$ than $A_{\rm AGN1}$, as noted by \citet{Delgado_2023}. The supernova parameters have a rather different effect, where increasing their values reduces the suppression. 
Again, this is consistent with the findings of \citet{Delgado_2023} who interpret this in terms of the nonlinear relationship between stellar and AGN feedback; strong stellar feedback suppresses the growth of black holes, weakening the effect of their feedback on matter clustering \citep{Pandey_2023,van_Daalen_2011}.

If we consider the Astrid model, we see that all of the feedback parameters appear in the parameter $\alpha_4$. $A_{\rm AGN1}$ appears linearly with a positive coefficient; $A_{\rm AGN2}$ also appears linearly but with a negative coefficient unless we have particularly weak supernova feedback ($A_{\rm SN2} < 0.56$); $A_{\rm SN1}$ again has a linear dependence but with a negative coefficient across our prior range; and $A_{\rm SN2}$ appears quadratically, with a coefficient which can take either sign, but is positive for the fiducial parameters. The only other places the feedback parameters appear is in $\alpha_7$ and $\alpha_8$ (which are in the denominator of the first term), where $A_{\rm AGN2}$ appears linearly with a positive coefficient, and $\alpha_1$ and $\alpha_3$, which are proportional to $A_{\rm SN 2}$. These effects combine to mean that, near the fiducial point, increasing $A_{\rm AGN1}$ will enhance the suppression of the matter power spectrum, whereas the opposite is true for $A_{\rm SN1}$ and $A_{\rm AGN2}$.
The impact of $A_{\rm SN2}$ depends on whether $\alpha_2 k$ is smaller than 1 or not. This is true for $k < 700 \, \hMpc$, and thus for all scales of interest. Hence, the $A_{\rm AGN2}$ parameter has two competing effects: its appearance in $\alpha_1$ and $\alpha_4$ seeks to increase the suppression as it increases, whereas its effect through $\alpha_3$ is to exponentially decrease this suppression. We find that this latter term is the dominant one, thus increasing $A_{\rm AGN2}$ decreases the suppression due to baryons.

For the IllustrisTNG model, we find that our expression does not include $A_{\rm AGN1}$, suggesting that the energy per unit black hole accretion rate for kinetic mode feedback does not significantly affect the matter power spectrum; upon repeating our analysis five times, only two of the six models depend on $A_{\rm AGN1}$.
This is consistent with \citet{Delgado_2023} who find little dependence of $S$ on this parameter. The parameter controlling the ejective speed and burstiness of this accretion does affect $S$, however, with $\alpha_5$ linear in $A_{\rm AGN2}$, and hence increasing $A_{\rm AGN2}$ increases the suppression. Similarly, $\alpha_5$ is linear in $A_{\rm SN1}$ (but with the opposite sign to $A_{\rm AGN2}$), which leads to a large variation in $S$ as the parameter is varied, where decreasing the feedback strength increases the power suppression, as we saw for the SIMBA model. 
Although $A_{\rm SN2}$ appears in two terms of our symbolic approximation ($\alpha_1$ and $\alpha_5$), one can combine these to see that $\log S$ is linear in $A_{\rm SN2}$. 
The sign of its coefficient is $k$-dependent, with it equal to zero at $k_\star =7.5 \, \hMpc$ at $z=0$ (for all parameters). At higher redshift its value depends on depends on $\Omega_{\rm m}$ and the transition occurs at increasingly large scales (smaller $k_\star$).
For $k < k_\star$, we find that increasing the value of $A_{\rm SN2}$ decreases the suppression of the matter power spectrum, whereas for larger $k$ the opposite is true. Similar $k$-dependent effects of the supernova parameters were also discussed in \citet{Delgado_2023}.

By inspecting the Swift-EAGLE model, we see that increasing any one of $A_{\rm AGN1}$, $A_{\rm AGN2}$ or $A_{\rm SN2}$ will increase the power suppression due to baryons.
The dependence on $A_{\rm SN1}$ is less straightforward, since it appears linearly in $\alpha_9$ (which has the effect of increasing suppression as $A_{\rm SN1}$ increases), however it appears in both the numerator and denominator of $\alpha_{10}$ and $\alpha_{11}$, so its overall affect on the power spectrum cannot be easily read off, and depends on the values of the other feedback parameters. Setting these to their fiducial values and considering $z=0$, we find that the dependence on $A_{\rm SN1}$ is not monotonic; for $k \lesssim 5 \, \hMpc$, increasing $A_{\rm SN1}$ initially decreases the amount of baryonic suppression, until we reach very large values ($A_{\rm SN1} \gtrsim 3$), where the suppression increases again. For smaller scales, the impact of $A_{\rm SN1}$ becomes almost monotonic, however now increasing $A_{\rm SN1}$ increases the amount of suppression.

Turning to the baryonification model, we see that almost all the feedback parameters appear in our model, except $\theta_{\rm out}$. When fitting this model to various hydrodynamic simulations, \citet{Arico_2021_baryons} note that this parameter is unconstrained, which is consistent with our findings. They also find that the parameters $M_1$ and $M_{\rm inn}$ cannot be inferred from these simulations. 
Evaluating the expression in \cref{tab:fits} using the fiducial parameters and $z=0$, we find that the term containing $\alpha_4$ is only relevant for $k \lesssim 10^{-3} \, \hMpc$, whereas the term containing  $\alpha_7$ becomes important only after $k \gtrsim 1.5 \, \hMpc$.
As such, the constraints on any feedback parameters which appear in these terms should be minimal, and indeed $M_1$ and $M_{\rm inn}$ only appear in $\alpha_4$ and $\alpha_7$. Similarly, $\theta_{\rm inn}$ also only appears in $\alpha_7$, and one observes that there is little variation among the simulations for this parameter. 
\citet{Arico_2021_baryons} also note that they infer $\eta \approx 0.5$ for all of the simulations considered. We can understand this, since for $10^{-3} \, \hMpc \lesssim k \lesssim 1.5 \, \hMpc$, one finds that the baryonification model is proportional to $k^{1+\alpha_6}$ (provided $\eta < 1$). Particularly for the lower end of this $k$ range, one would expect that most of the models should behave similarly, and thus (given $\alpha_6$ only depends on $\eta$) it is unsurprising given our fit that all ``reasonable'' simulations require a similar value of $\eta$.

\subsection{Small-scale extrapolation behaviour}

When obtaining the emulators listed in \cref{tab:fits}, we were limited in terms of the maximum $k$ value we could use. Although we enforced given extrapolation behaviour for large $z$ or low $k$, the small-scale (high $k$) extrapolation behaviour was left completely free. In this section we investigate how the models extrapolate as one considers this limit.

Beginning with Astrid, we find that the first term in \cref{tab:fits} is proportional to $k^{\alpha_3 -1}$ for large $k$, whereas the second term is proportional to $k$. Given the form of $\alpha_3$, we find that the first term is dominant at large $k$ if $A_{\rm SN 2}>2.1$, but this is outside the prior range in \cref{tab:hydro_prior}. As such, $\log S \propto k$ at large $k$ with a positive coefficient of proportionality, and thus $S$ grows exponentially on small scales.

We find different behaviour for both IllustrisTNG and Swift-EAGLE, for which the first terms are proportional to $k^\gamma \exp(-\lambda k)$ at large $k$, where $\gamma = 2$ for IllustrisTNG and $\gamma=1$ for Swift-EAGLE. For both cases, $\lambda > 0$ and thus at very large $k$, this term disappears, and the offset term dominates. These both tend to constants at large $k$, although this constant is independent of $z$ for IllustrisTNG, whereas it grows with $z$ for Swift-EAGLE.
These behaviours seem unlikely on physical grounds (one may expect that $S$ would continue to grow on small sclaes) and are thus extrapolation artefacts of the fitting functions. However, when evaluated at the fiducial parameters, the growth with $k$ does not begin to flatten out until $k \sim 50 \hMpc$ for IllustrisTNG and $k \sim 10^3 \hMpc$ for Swift-EAGLE, so for $k$ values far beyond the range of the training data.

The extrapolation properties of SIMBA is a function of redshift. For all $z$, the first term in \cref{tab:fits} is proportional to $k^{2-\alpha_4 z - \alpha_6}$ with a negative coefficient, hence, at low redshift ($z \lesssim 0.5$) $\log S \to -\infty$ as $k \to \infty$, thus $S \to 0$ in this limit. This is surprising given the growth with $k$ seen in \cref{fig:example_simulations}, however we find that the maximum in $\log S$ is not reached until $k \sim 10^{3.5} \hMpc$ for the fiducial parameters, and thus this is an artefact which is not reached until we extrapolate by several orders of magnitude in $k$. For higher redshift, the first term tends to 0 at high $k$, and hence we find that $S$ tends to a $z$-dependent constant at large $k$.

The baryonification model has a more physical extrapolation behaviour, with $\log S$ proportional to $k^{2+\alpha_6}$ in this regime. $\alpha_6$ is only dependent on the parameter $\log_{10}(\eta)$, and its range in \cref{tab:baryonification_prior} means that the exponent is always positive. The sign of the proportionality constant is parameter dependent (and is equal to the negative of the sign of $\alpha_7$), but in all but the most extreme cases it is positive, meaning that $S$ almost always grows as $k^{2+\alpha_6}$ on small scales, but it can also decay for appropriately chosen parameters.

\subsection{Comparison to previous work}

Several other works have attempted to find symbolic approximations to baryonic suppression on the matter power spectrum. For example, \citet{Gebhardt_2024} use symbolic regression to approximate this quantity for the SIMBA simulation as
\begin{equation}
    \label{eq:Gebhardt_model}
    S - 1 = a_0 \left( k \left( a_1 \mathcal{S} - a_2\right) \right)^{1/4} - a_3,
\end{equation}
where $a = \{ \splitatcommas{0.25, 0.35, 1.09, 0.14} \}$ and $\mathcal{S}$ is the median normalised gas spread metric, which is a proxy for how far a gas particle is displaced due to the presence of baryonic effects.
Other metrics have been used to parametrise the effect of baryons when making such predictions; for example, \citet{van_Daalen_2020} choose to provide a fitting function for the matter power suppression in terms of the baryonic fraction of massive halos.
We note that \citet{Gebhardt_2024} use the 1P CAMELS parameter set, which varies one parameter at a time with the others set to their fiducial values, whereas we explore the full prior of \cref{tab:hydro_prior} by using the CV set. Although this model is extremely simple and roughly obeys the correct trends when evaluated on the CV set, the error distribution in Figure 11 of \citet{Gebhardt_2024} is significantly larger than those given in \cref{fig:error_on_S}, hence our models are much more accurate. Moreover, \cref{eq:Gebhardt_model} does not obey the correct analytic limit as $k \to 0$, and hence violates the causality constraint on large scales.

An alternative parametrisation which was introduced by \citet{Amon_2022} supposes that the nonlinear power spectrum in the presence of baryons, $P(k, z)$, can be expressed in terms of the linear and nonlinear dark-matter-only power spectra, $P_{\rm nbody}^{\rm L}(k,z)$ and $P_{\rm nbody}(k,z)$, respectively. One writes
\begin{equation}
    P (k, z) = P_{\rm nbody}^{\rm L}(k,z) + A_{\rm mod}(k,z) \left(P_{\rm nbody}(k,z) -  P_{\rm nbody}^{\rm L}(k,z)\right),
\end{equation}
where $A_{\rm mod}$ is an amplitude parameter which is assumed to be independent of $k$, $z$ and background cosmology. 
\citet{Preston_2023} extended this analysis by binning $A_{\rm mod}$ in $k$ bins, such that it can depend on wavenumber, but not redshift or cosmology.
The approach is similar in spirit to \citet{Bielefeld}, who suggest adding extra fitting parameters to non-parametrically fit $S(k,z)$ within a cosmological analysis.
\citet{Schaller_2025} modify this further by assigning an explicit functional form to $A_{\rm mod}(k)$, for which they choose a sigmoid function
which depends on three parameters:
$A_{\rm low}$, $k_{\rm mid}$ and $\sigma_{\rm mid}$. Like our fitting functions, this obeys the correct analytic limit as $k\to 0$. However, when fitted against the FLAMINGO simulations \citep{Kugel_2023,Schaye_2023}, this sigmoid fit does not produce an upturn in the power spectrum ratio for $k \gtrsim 10 \, \hMpc$, but is able to fit the data well before this point. 
Our models can capture this feature (see, e.g., \cref{fig:example_simulations}), thus representing an advantage of our formulae. 
\citet{Schaller_2025} find that $A_{\rm low}$ and $\sigma_{\rm mid}$ are almost constant across their simulation suite, suggesting that only one parameter is required.
Future work could be dedicated to seeing whether this model also fits the CAMELS simulations' power spectra and how the parameters of this model depend on the varying feedback parameters.

The most similar analysis to our study is that of \citet{Sharma_2024}, who model $S$ for the same hydrodynamic simulations are this work, however they use a Gaussian Process rather than symbolic fits. As described in \cref{sec:results}, we obtain similar accuracy to \citet{Sharma_2024}, suggesting that, despite their highly flexible model, the simple formulae given in \cref{tab:fits} are sufficient for accurately describing this phenomenon. 

\citet{Schaller_2025_Flamingo} also train a Gaussian Process emulator for this task -- on the FLAMINGO suite of simulations rather than CAMELS -- and are able to obtain high accuracy predictions of 1\% error up to $k = 10 \, \hMpc$. We have not considered these simulations in this work, but they could represent an ideal target for future SR analyses. Interestingly, they find that $S$ only converges on small scales once the box size is approximately $200 \, \Mpch$ per side, which is significantly larger than CAMELS. Although we account for this sample variance using the CV set and obtain models for $\varepsilon(k,z)$, this suggests that for a cosmological analysis, one could reduce the magnitude of $\varepsilon(k,z)$ in the likelihood, since 
only the error in the fit (and not the sample variance) should be significant for larger volumes.

One need not parametrise the effect of baryons in terms of the input hydrodynamic and cosmological parameters, but instead use emergent properties of the simulation. For example, the $SP(k)$ emulator of  \citet{Salcido_2023} is able to reproduce the matter power spectrum accurately from the mean baryon fraction of halos.
Similarly, \citet{Pandey_2023} also study power suppression in the context of the CAMELS simulations, choosing to use a random forest to predict power suppression from observable quantities. The uncertainties on their model (see Figures 6 and 7 of \citet{Pandey_2023}) are slightly larger than the estimate from the CV set, justifying our claim in \cref{sec:results} that it is not a concern that the normalised errors in \cref{fig:norm_error_on_S} extend to greater than unity. 
\citet{Pandey_2023} also show in their appendix that, although the $N$-body power suppression is relatively robust to the choice of box size for the $k$ range we consider, it is somewhat sensitive to the resolution. This again supports our decision to not attempt to try to further reduce the errors from \cref{fig:norm_error_on_S}, since there is an additional systematic error on $S$ which has not been considered here.
\citet{Delgado_2023} also train a random forest on the CAMELS simulations to make this prediction but as a function of other variables. An extensive comparison of their conclusions to our models is given in \cref{sec:parameter_dependence}.

\section{Conclusion}
\label{sec:conclusion}

To fully exploit data from current and upcoming cosmological surveys, we must be able to model the non-negligible effects of baryons on the clustering of matter in the Universe. Different hydrodynamical models yield significantly different (and highly stochastic) predictions for the impact of baryons. These simulations are computationally expensive, and hence emulating their behaviour is essential if one wishes to incorporate these within cosmological analysis pipelines. 

As such, in this work we used symbolic regression to obtain symbolic approximations for the suppression of the matter power spectrum due to the inclusion of baryons for four implementations of sub-grid physics in hydrodynamical simulations (Astrid, IllustrisTNG, SIMBA and Swift-EAGLE) as well as one baryonification model. 
This suppression is sufficient for modelling baryonic effects at the field level for scales up to $k \sim 10 \, \hMpc$ \citep{Sharma_2024}.
These expressions are given in \cref{tab:fits}, which summarises the main results of the paper.
For the hydrodynamical simulations, our models achieve similar accuracy to previous (non-symbolic) emulators, and we characterise this error -- which is comparable to our estimate of sample variance -- also in terms of symbolic expressions (\cref{tab:sigma_fits}). 
Our fits for the baryonification model have a root mean squared error of 0.7\% compared to the \bacco{} emulator \citep{Arico_2021_baryons}, although this error increases towards small scales and must be combined with the error on the \bacco{} emulator itself compared to hydrodynamic simulations (approximately 1-2\%).

By construction, our models extrapolate to give the correct physical behaviour on the largest scales and at early times. Since they are symbolic, we are able to easily determine the impact of varying cosmological and hydrodynamical parameters to gain physical insight into how feedback processes affect the clustering of matter. Our models are compact and are comprised of elementary mathematical operations, and are thus trivial to export to programming language of the user's choice without reliance on external packages (although we make python implementations publicly available).

We chose to focus on the publicly available CAMELS simulations and one baryonification model, however future work should be dedicated to applying a similar methodology to a broader range of models, to enable their comparison at the level of the power spectrum (for example, using the power spectrum library of \citet{van_Daalen_2020}).
One of the advantages of the baryonification model is its flexibility such that, for appropriate choices of parameters, it can approximate many different hydrodynamic simulations \citep{Arico_2021_baryons}. In future work, it would be interesting to investigate the flexibility of the other models in \cref{tab:fits} and whether these models are distinguishable when used in cosmological analyses.
For the hydrodynamical simulations, one could also extend the number and range of parameters varied to see what impact other feedback parameters have on the power spectrum. In terms of cosmological parameters, it would be particularly interesting to vary $\Omega_{\rm b}$ alongside $\Omega_{\rm m}$ and $\sigma_8$ in the hydrodynamical simulations, since these are the only three cosmological parameters which are important for determining the suppression within the baryonification model studied here.
Finally, given the substantial sample variance observed between different CAMELS simulations, it would be useful to investigate how our models behave when applied to larger volume simulations, where this effect should be smaller, for example with the FLAMINGO simulations \citep{Kugel_2023,Schaye_2023}. 
One could also extend our emulation approach to quantity the effects of baryons on other quantities of interest in cosmology, such as the matter bispectrum.

Our approach has been centred on the idea of conditioning on a particular model (i.e. find a single symbolic approximation for each model) rather than obtaining a single, flexible model which is robust to the implementation of hydrodynamic physics. This should enable future work to embed these different models within pipelines for inference on real cosmological datasets, and thus one can compute evidence ratios to determine which model is favoured by the data, and thus inform the development of future subgrid models.

In summary, we have demonstrated that symbolic regression offers a powerful, interpretable, and highly portable method to emulate the impact of baryonic physics on cosmological quantities such as the matter power spectrum. By providing compact analytic forms tailored to specific hydrodynamical and baryonification models, our results bridge the gap between high-fidelity simulations and practical cosmological inference. These symbolic models not only reproduce the known behaviour of baryonic suppression with competitive accuracy but also offer new opportunities for model comparison and parameter exploration with upcoming surveys.

\section*{Acknowledgements}

We thank David Alonso, Carlos Garcia-Garcia, Francisco Villaescusa-Navarro and Matteo Zennaro for useful comments and suggestions. 
DJB was supported by the Simons Collaboration on ``Learning the Universe'' and is supported by Schmidt Sciences through The Eric and Wendy Schmidt AI in Science Fellowship.
HD is supported by a Royal Society University Research Fellowship (grant no. 211046).
PGF acknowledges support from STFC and the Beecroft Trust. 
The data underlying this article will be shared on reasonable request to the corresponding authors.
We provide python implementations of the models in \cref{tab:fits,tab:sigma_fits} at \url{https://github.com/DeaglanBartlett/symbolic_pofk}. 

\bibliographystyle{aa}
\bibliography{references}

\begin{appendix} 

\section{Pareto Fronts}

\begin{figure*}[t]
    \centering
    \input{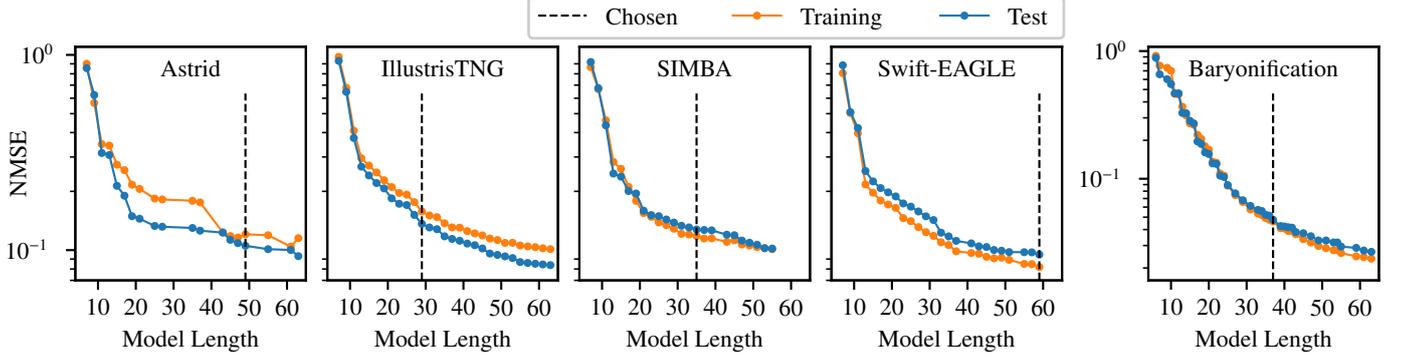}
    \caption{
    Pareto front of solutions found with \operon{} for $\log S$ (\cref{eq:S_definition}).
    We plot the training and test sets separately, and the dashed lines indicates our chosen models.
    } 
    \label{fig:pareto}
\end{figure*}

In this appendix we plot the Pareto fronts for the models of $\log S$. These are given in \cref{fig:pareto}, where we plot the training and test losses (NMSE) against model length. We choose the models indicated by the dashed vertical lines, using the selection criteria outlined in \cref{sec:symbolic_regression}, and these are given in \cref{tab:fits}.

\section{Comparison of emulator error to sample variance}

\begin{figure*}[b]
\input{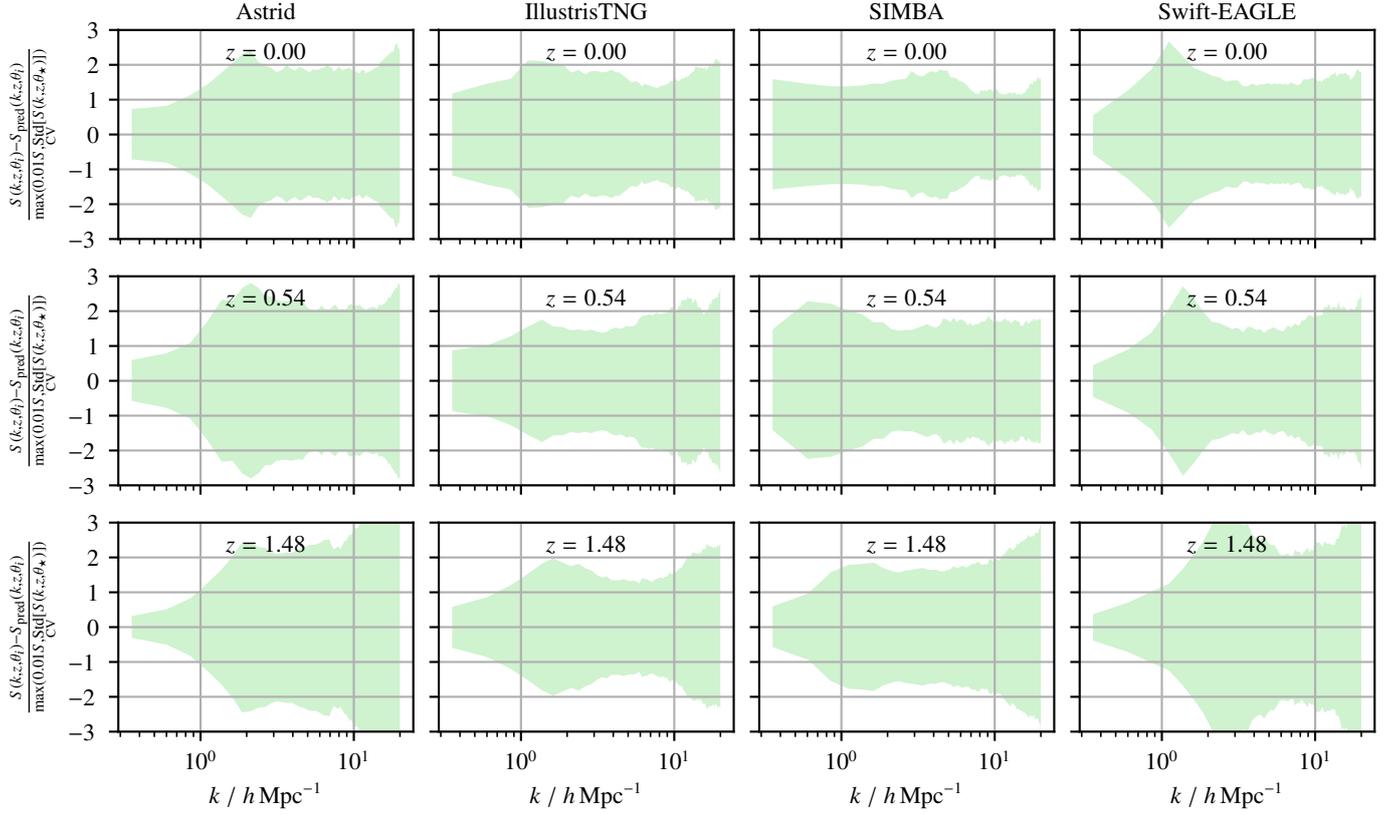}
\caption{
Same as \cref{fig:error_on_S}, but where we divide the difference between the truth and prediction by the standard deviation of the CV set. For a perfect symbolic fit and if this standard deviation perfectly captured the stochasticity, then the shaded region would be a band between -1 and 1.
}
\label{fig:norm_error_on_S}
\end{figure*}

This appendix compares the accuracy of our symbolic emulators for the hydrodynamical simulations (evaluated on the test set) to the sample variance given by the CV set. The ratios of these two quantities are given in \cref{fig:norm_error_on_S}. If our emulators were perfect and the CV set exactly captured the sample variance, then the bands plotted in \cref{fig:norm_error_on_S} would be horizontal and span between $-1$ and 1. This is not quite true, with the bands typically at 1.5 to 2$\sigma$. As discussed in \cref{sec:results}, we expect that this is dominated by the inaccuracies of the sample variance estimation.

\end{appendix}

\end{document}